\documentclass[authoryear, preprint,10pt]{elsarticle}

\usepackage{amssymb}
\usepackage{xcolor}
\usepackage{graphicx}
\usepackage{import}
\usepackage{rotating}
\usepackage{lscape}
\usepackage{pdflscape}
\usepackage{caption}
\newcommand\fnote[1]{\captionsetup{font=scriptsize}\caption*{#1}}

\usepackage[hyphens]{url}
\usepackage[hidelinks]{hyperref}
\hypersetup{breaklinks=true}
\usepackage{cite}
\usepackage{setspace}
\usepackage{etoolbox}
\usepackage{svg}
\usepackage{amsmath}

\usepackage[lmargin=20mm,rmargin=20mm,tmargin=20mm,bmargin=15mm, footskip=0.7cm]{geometry}

\BeforeBeginEnvironment{equation}{\begin{singlespace}}
\AfterEndEnvironment{equation}{\end{singlespace}\noindent\ignorespaces}

\journal{Finance Research Letters}

\begin{document}

\begin{frontmatter}



\title{The role of investor attention in global asset price variation during the invasion of Ukraine}


\author[inst1]{Martina Halousková}
\author[inst1]{Daniel Stašek}

\affiliation[inst1]{organization={Department of Finance, The Faculty of Economics and Administration, Masaryk University},
            addressline={Lipova 41a}, 
            city={Brno},
            postcode={60200}, 
            country={Czech Republic}}

\author[inst1]{Matúš Horváth}


\begin{abstract}

We study the impact of event-specific attention indices -- based on
Google Trends -- in predictive price variation models before
and during the Russian invasion of Ukraine in February 2022. We extend
our analyses to the importance of geographical proximity and economic
openness to Russia within 51 global equity markets. 
Our results demonstrate that 36 countries show significant attention to the conflict at the onset of and during the invasion, which helps predict volatility.
We find that the impact of  attention  is  more  
significant  in  countries  with a higher degree of economic openness 
to Russia and those nearer to it.

\end{abstract}




\end{frontmatter}


\section{Introduction}
\label{sec:intro}

\begin{figure}[ht]
\centering
\caption{Conflict attention index and daily price volatility on four continents}
\label{fig:story-chart}
\includegraphics[width=12.5cm]{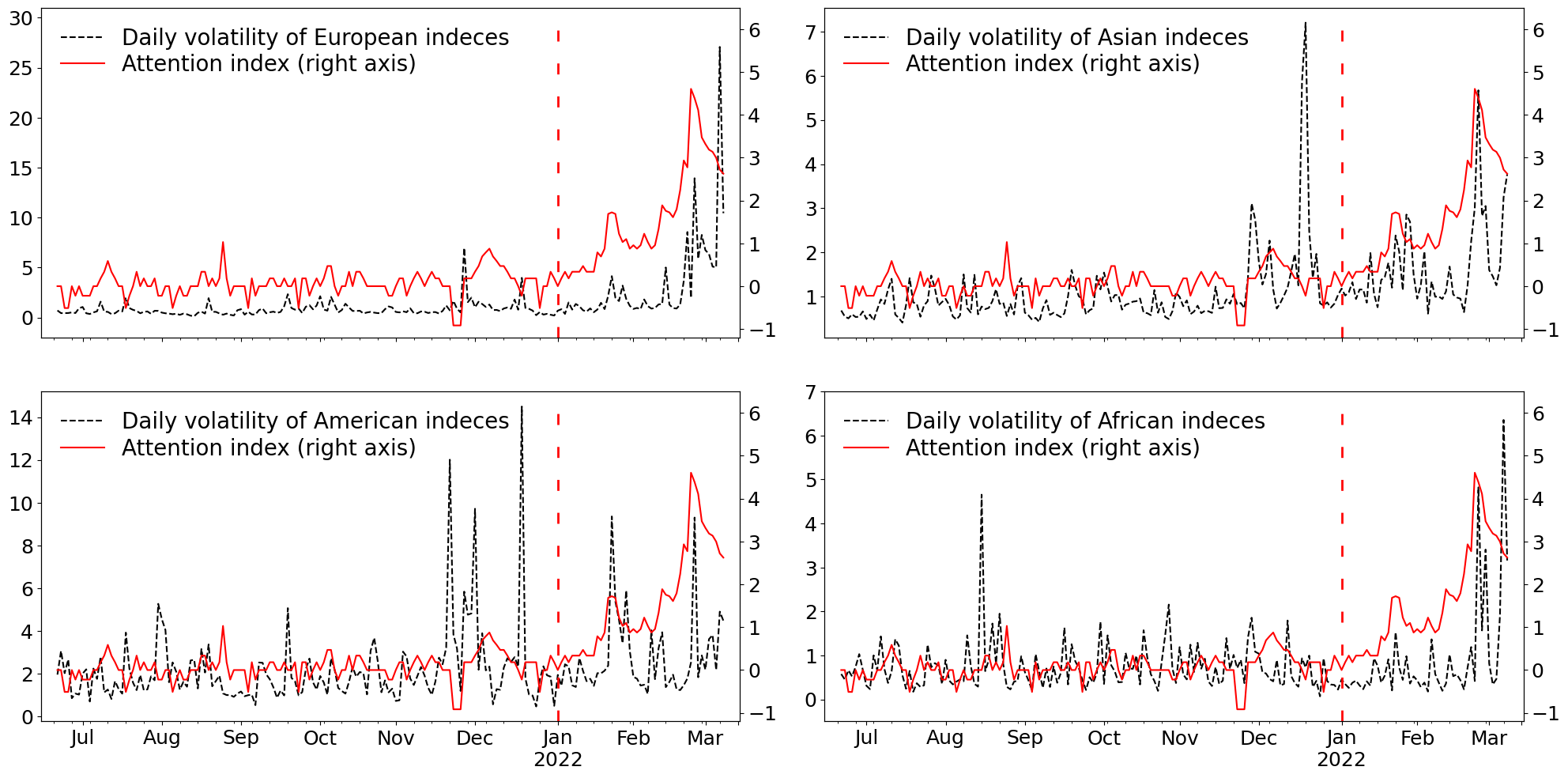}
\fnote{Notes: The daily volatility for each area is computed as an
  average of the individual volatilities of available Morgan Stanley Capital International (MSCI) indices for
  the particular area. Furthermore, for the Americas volatility
  average, we used both the North and South American MSCI indices. The
  vertical red dashed line divides the time series into two periods. We label them the pre-invasion and onset-of-invasion periods.}
\end{figure}

The prelude to the Russo--Ukrainian war began in mid-October 2021,
during which Russian forces gathered near Ukraine's borders and in the
occupied Crimea region \citep{CNN}. Shortly thereafter, aggressive and
escalating statements from Russian policymakers were reported
\citep{Bowen2022}, which led Joe Biden to announce consequences in the
event of any Russian invasion of Ukraine \citep{Reuters1}. As a result
(on December 8), we observe the first sharp growth in the conflict
attention index depicted in Figure \ref{fig:story-chart}, meaning that
this threat was likely recognized by many. The intelligence provided
by Western security agencies suggested that a possible Russian
invasion could start in early 2022 \citep{WP}. Considering those
warnings, combined with the growth of tensions, military movements
\citep{DJIN}, and accusations \citep{FT}, the attention paid to the possible conflict grew rapidly. 

In his speech on February 21, 2022, President Putin announced that
Russia had recognized the separatist republics in eastern Ukraine,
which was followed by consequent threats to Ukraine
\citep{Reuters2}. In the early morning of February 24, Russian armed forces began a full-scale war against Ukraine.

The response was to impose unprecedented economic sanctions on the
Russian economy, which were implemented shortly after the invasion and
targeted all kinds of industries such as banking, oil exports, and
high-tech components. However, due to economic interconnectedness and
dependency on Russian commodities, those sanctions also created risks
for all companies and households in countries economically linked to
Russia. The sanctions are already resulting in recognizable economic consequences, especially for Europe, whose largest energy supplier is Russia (in 2021, approximately 45\% of imported natural gas came from Russia \citep{IEA}). Along with a sharp increase in energy prices, we are also witnessing devastating consequences for countries that rely on Ukrainian wheat imports \citep{Wheat}. 
Thus, this war has evolved into a serious global economic and
political issue that is subject to exceptional worldwide attention.

In this context, our aim is to investigate this extraordinary interest and determine whether it is linked to increased volatility of stock markets around the globe. We build upon a growing literature related to the relationship between future market movements and investor behavior, particularly investor sentiment and investor attention.

From a theoretical perspective, we rely on the limited attention
hypothesis of \citet{Barber2008}, according to which investors face a
difficult task of choosing among numerous investment opportunities, despite possessing limited time and resources, and thus gravitate toward "attention-grabbing" options. 
\citet{Andrei2015} expand on this idea to demonstrate that news that receives substantial interest takes less time to be incorporated into prices. Their results also suggest that investors seek more information during times of high uncertainty -- such as the unexpected Russian military invasion of Ukraine. 

To capture investors' panic, fear, and uncertainty, we use a very
popular direct measure of attention, the Google Search Volume Index
(SVI). Some of the first applications of Google query data in research
came from epidemiology \citep{Ginsberg2009, Dugas2013}; however, such
approaches rapidly spread to the field of finance \citep{DA2011,
  Joseph2011} as a proxy for attention (albeit sometimes imprecisely
referred to as a sentiment proxy). 
Search volume has been shown to be correlated with lagged trading volumes \citep{Preis2010, Bordino2012}, to improve trading strategies \citep{Preis2013, Bijl2016} or diversification strategies \citep{Kristoufek2013}. For economic mechanisms explaining how the attention may impact future volatility, refer, for example, to \citet{Aouadi2013, Vlastakis2012, Hamid2015, Dimpfl2016, Audrino2020}. 
Several studies have also examined interest in cryptocurrencies, measured with Google, as a driver of their price and volume fluctuations \citep{Kristoufek2013, Kristoufek2015, Garcia2014, Cheah2015, Cretarola2017, Urquhart2018, Aalborg2019, Eom2019, Burggraf2020, Chen2020}. Similar results can be found for major FX markets \citep{Kita2012, Smith2012, Goddard2015, Han2018, Wu2019, Saxena2020, Kapounek2021}.

Investor attention appears to be particularly effective in studies
related to specific events, which is also the case for our study,
namely attention devoted to macroeconomic developments
\citep{Lyocsa2020, Plihal2021}, earnings news announcements
\citep{Hirshleifer2011, Hirshleifer2021, Fricke2014GoogleSI,
  Ben-Rephael2017}, the outbreak of the COVID-19 pandemic \citep{Chen2020, Lyocsa2020a}, and even this military conflict \citep{Lyocsa2022}. 

We contribute to this literature on the impact of event-specific
attention by exploring the one-day ahead predictive power of investor attention
devoted to the military conflict in Ukraine in volatility models. We
divide the data into two samples (the pre-invasion and
onset-of-invasion periods) to compare the effects of such attention on
price fluctuations.\footnote{We use January, 1 2022, as the date to
divide the data based on an undisclosed U.S. intelligence report
warning of a Russian invasion of Ukraine in early 2022, first
published in the Washington Post on December 3, 2021
\citep{WP}. Thus our second sample captures the period of heightened
attention and risk of an upcoming invasion.} The analysis covers the stock indices of 51 countries. Our goal is to compare these results and determine whether (1) geographical or (2) economic proximity to the conflict influences the impact of conflict attention on volatility. 

The remainder of this paper is organized as follows. In section
\ref{sec:datamethodology}, we describe the data, their sources, and
how they were processed into the measures used for analysis. In the
following section \ref{sec:results}, we present and describe the
results and emphasize their place in the context of the war and
economic connectedness. Finally, we conclude by describing our results and contributions.

\section{Data and methodology}
\label{sec:datamethodology}

\subsection{Financial data}

Our financial data consist of two datasets -- the daily prices of
selected indices and economic indicators. The first dataset includes
data from all countries with an available MSCI index. MSCI indices
were selected because of the consistent methodology used to calculate
indices for all included countries. In addition to the MSCI indices,
we include the Latvian stock index -- OMX Riga -- to capture the
impact in this Baltic state. The data were collected from a Bloomberg
terminal as an OHLC\footnote{An abbreviation for open, high, low,
  close data.} dataset. After removing countries with missing OHL
observations, our sample covers 51 countries, including 8 regions in
the Americas, 3 regions in Africa, 19 European countries, and 21
Asia-Pacific regions, including Australia. The period of study is from
June 22, 2021, to March 8, 2022. For modeling purposes, the dataset
was divided into two periods -- before the invasion of Ukraine, June
22, 2021, to December 31, 2021, and during the onset of the invasion
of Ukraine from January 1, 2022, to March 8, 2022. The median number
of observations in the pre-invasion and onset-of-invasion samples is 136 and 47, respectively. 

The second financial dataset considers a filtered set of countries based on previously mentioned conditions. 
For each country, we extracted data on imports from Russia and exports to Russia, as well as GDP, all for the year 2020.
The source of these data is the UN Comtrade Database. This dataset was
used to calculate the degree of openness (DOO)
\citep{rodriguez2000degree} to Russia of country $i$ country as follows: 
\begin{equation}
    DOO_i=\frac{Export_{i} + Import_{i}}{GDP_i},
\end{equation}
where $GDP_i$ is the GDP of country $i$, $Export_{i}$ is the exports
of country $i$ to Russia and $Import_{i}$ is the imports of country $i$ from Russia. In addition, we define a variable $Dist_i$, as a distance to Moscow from $ith$ country capital. The distance is measured in the order of $10^3 km$.

\subsection{Attention measures}

Our attention measures were retrieved from Google Trends with the help
of the R package gtrendsR \citep{gtrendsR}. 
Unfortunately, the
availability of daily data is limited to 270-day intervals, and longer
samples would require additional scaling. Although this may appear to be a relatively short sample, we opt for the 270-day interval, as it sufficiently covers the events we want to consider. Previous literature points to the presence of seasonality in attention. Thus, if our sample was larger, we could use fixed effects to address this issue \citep{liu2021}.

We use two sets of search terms to construct two variables, one
related to general stock market attention and one for the attention
paid to the military conflict, denoted $G_t$ and $C_t$,
respectively. We include the general index to ensure that we measure
the effect of excess attention to the conflict adjusted for the
general day-to-day interest of investors in trading. Since we sought
to capture global interest, we opted for queries of topics -- an
option that automatically translates keywords into all available
languages and accounts for spelling variations. 
We primarily wanted to aim our research at international investors, 
as the developed financial markets are very tied to each other, 
reflecting the global macroeconomic shocks and unprecedented events 
such as this one. Thus we decided to retrieve the data using the "worldwide" option. 
The $G_t$ and $C_t$ indices are then adjusted for the time zones of the stock exchanges
corresponding to each MSCI index in our dataset. For the MSCI indices
that cover more than one country, we select the time zone with the
majority coverage, as reported in the country weights in the MSCI fact sheets. Table \ref{C-G-Descriptives} provides summary statistics for indices $G_t$ and $C_t$ after log transformation in three different time zones. 

After accounting for time zones, we also remove values for nontrading
days. Our approach consists of taking the maximum value of Friday to
Sunday and assigning this value to Friday, and the method is applied for holidays. This procedure must be applied individually to each country, as the nontrading days are not identical. 
\begin{table}[h!]

\centering
\renewcommand{\arraystretch}{1.1} 
\small

\caption{Descriptive statistics of variables of conflict and general attention indeces}
\label{C-G-Descriptives}
\begin{center}

\begin{tabular}{ l c c c c c c c } 

\hline
 & Mean & S.D. & Median & Min. & Max. & $\rho(1)$ & $\rho(5)$ \\
\hline
 $C_t$ (UTC+0) & 0.453 & 0.936 & 0.182 & -0.916 & 4.605 & 0.944 & 0.796 \\
 $C_t$ (UTC+6) & 0.549 & 0.912 & 0.223 & -0.693 & 4.605 & 0.940 & 0.809 \\
 $C_t$ (UTC-6) & 0.452 & 0.937 & 0.182 & -0.916 & 4.605 & 0.943 & 0.797 \\
\hline
 $G_t$ (UTC+0) & 3.862 & 0.147 & 3.861 & 3.188 & 4.178 & 0.609 & 0.410 \\
 $G_t$ (UTC+6) & 3.862 & 0.149 & 3.861 & 3.188 & 4.178 & 0.630 & 0.458 \\
 $G_t$ (UTC-6) & 3.860 & 0.148 & 3.856 & 3.188 & 4.177 & 0.629 & 0.430 \\
\hline

\end{tabular}
\end{center}

\textit{Notes: S.D. stands for standard deviation, $\rho(1)$
  represents the first-order autocorrelation coefficient and $\rho(5)$
  denotes the fifth-order autocorrelation.}
\end{table}

We used five topic search terms to create the conflict index $C_t$
('Russia', 'Ukraine', 'Vladimir Putin', 'NATO', and 'sanctions') and
31 topics \ref{sec:appendix} 
to construct the general attention index $G_t$. We based our general topic selection on previous studies that captured general attention via a list of financial keywords. In particular, our list of topics closely resembles keywords used by \citet{Mao2011, Preis2013}. In addition, our initial list of search words was first reduced to those keywords that could be obtained as Google topics. Furthermore, to ensure that the resulting average $G_{t}$ variable captures one common phenomenon, we prune the list of keywords by removing topics whose SVI is negatively correlated with the rest of the topics. From previous experience working with attention measures in volatility forecasting, this usually helps remove noise from the data.

Both attention indices are then calculated as simple averages of the individual variables. 
These data take the form of a normalized volume ratio on a scale of 0--100, where 100 represents the maximum search activity during the selected period. 
The acquired search volume ratio $SVI_t$ at time $t$ can be further
transformed according to the procedure proposed by \citet{DA2011} into
the abnormal search volume index ($ASVI_{t}$), which should help us to identify significant changes in Google searches.

The $ASVI_{t}$ is usually applied to address the noisiness of $SVI_t$ and to capture only abnormal search activity. However, it does not capture the scale of the abnormal change. 
As the data we are working with have a rather unusual shape, with nearly exponential growth around the time of the invasion of Ukraine, we concluded that the $ASVI_{t}$ transformation is not suitable for our analysis. Instead, we opt for a log transformation of $SVI_t$. 
For the remainder of the paper we use $G_t$ and $C_t$ in this log-transformed form.

\subsection{Price variation estimator}

As we rely on MSCI indices, we are limited to the use of daily OHLC
data, which do not allow us to use the standard realized volatility
estimator calculated as the sum of squared intraday returns (see,
e.g., \citet{andersen2003modeling}). Thus, to maintain the stylized
facts\footnote{The realized range-based estimator defined in Equation \ref{eq:vol_estimator} maintains the features of the volatility time series for most of the MSCI indices data and, most important, its  persistency. For more information, see Table \ref{daily-vol-descriptives}. An alternative would be to model price ranges, for example as in \citet{baig2019price}. } of the volatility time series, we employ a range-based volatility estimator following the approach of \citet{lyocsa2021yolo}, which was originally motivated by the approach of \citet{patton2009optimal}, who notes that since the true data generating process is unknown, the optimal estimator must also be unknown. Therefore, a combination of several estimators may be less prone to estimator choice uncertainty:

\begin{equation}
\label{eq:vol_estimator}
V_t = 100^2\times(J_t+3^{-1}(PK_t+GK_t+RS_t))
\end{equation}
\\
where $ V_t $ is the price variation estimator. Furthermore, for every
day $ t = 1, 2, ..., T $, we compute the average of three different
realized range-based estimators ($ PK_t $, $ GK_t $, $ RS_t $) and
adjust the final price variation estimates for overnight price
variation $ J_t $. We denote by $ PK_t $ the estimator of \citet{parkinson1980extreme}:

\begin{equation}
PK_t = \frac{(h_t-l_t)^2}{4\ln2}
\end{equation}
\\
by $ GK_t $ the volatility estimator of \citet{garman1980estimation}
is defined as:

\begin{equation}
GK_t = 0.511(h_t-l_t)^2-0.019(c_t(h_t+l_t)-2h_tl_t)-0.383c_t^2
\end{equation}
\\
by $ RS_t $ the \citet{rogers1991estimating} estimator takes the form:

\begin{equation}
RS_t = h_t(h_t-c_t)+l_t(l_t-c_t)
\end{equation}
\\
and finally, the overnight price variation is defined as:

\begin{equation}
J_t = [\ln(Open_t)-\ln(Close_{t-1})]^2
\end{equation}
\\
where $ Open_t $, $ High_t $, $ Low_t $, and $ Close_t $ represent the open, high,
low and close prices on a given day $ t $. Furthermore, define $ h_t = \ln(High_t) - \ln(Open_t) $, $ l_t = \ln(Low_t) - \ln(Open_t) $, $ c_t = \ln(Close_t) - \ln(Open_t) $.  

The estimated daily volatilities are in \ref{daily-vol-descriptives}.

\subsection{Model specification}

With a focus on the attention variables and subsequent exploration of
their impact on price variation, we define a parsimonious model, mimicking the well-known and time-tested heterogeneous autoregressive (HAR-RV) model of \citet{corsi2009simple}. By doing so, we investigate whether the attention 
to the invasion of Ukraine prompted individual investors to increase their trading activities, which would raise volatility.
In contrast to the standard HAR-RV, we omit the monthly component
because, as presented in Table \ref{daily-vol-descriptives}, the fifth-order autocorrelation is low in some cases.
During our main period of interest, the uncertainty about subsequent
price developments is so high that what the last month's price
variation was should rarely matter. 

\begin{equation}
\label{eq:har-model}
V_{t+1} = \beta_0 + \beta_1 V_t + \beta_2 V_t^w + \beta_3 C_t + \beta_4 G_t + \epsilon_{t+1}
\end{equation}
\\
where $ C_t $ and $ G_t $ are the attention variables defined in the
previous section. $ V_t^w $ is the weekly price variation component
given as $ V_t^w = 5^{-1}\sum_{j=0}^4 V_{t-j} $. The model in Equation
\ref{eq:har-model} is estimated via ordinary least squares and for
both data samples, that is, in the pre-invasion and onset-of-invasion
periods. Next, we estimate the model with the log-transformed price variation.

In order to test the significance of attention variables across the worldwide indices and also hypotheses developed further in the results section, we opted to utilize panel data regression models, defined as:

\begin{equation}
\label{eq:panel-har}
    V_{i, t+1} = \beta_0 + \beta_1 V_{i, t} + \beta_2 V_{i, t}^w + \beta_3 C_{i,t} + \beta_4 G_{i, t} + \alpha_i + \epsilon_{i, t+1}
\end{equation}

\begin{equation}
\label{eq:panel-har-ext}
    V_{i, t+1} = \beta_0 + \beta_1 V_{i, t} + \beta_2 V_{i, t}^w + \beta_3 C_{i,t} + \beta_4 G_{i, t} + \beta_5 C_{i,t}X_i + \alpha_i + \epsilon_{i, t+1}
\end{equation}
\\
where e.g. $V_{i, t}$ is volatility of $ith$ cross-sectional unit at time $t$, $\alpha_i$ is the unobserved heterogeneity and $X_i$ stands for either the $DOO_i$ or $Dist_i$. We estimate the model parameters via fixed effect panel data regression with standard error of \citet{driscoll1998consistent} robust to cross-sectional and temporal dependence. Hence, we name the Equation \ref{eq:panel-har} as FE-DK model and Equation \ref{eq:panel-har-ext} FE-DK-DOO and FE-DK-Dist.

\section{Results}
\label{sec:results}

Firstly, we explore the significance of the independent variables across all 51 countries via estimating Equation \ref{eq:panel-har}. As presented in Table \ref{table:prelim-panels}, in the pre-invasion period all regressors, except the conflict attention variable are significant, which is to be expected. In contrast, within the onset-of-invasion period, the $C_t$ is among the most significant, explaining high portion of the model variability. This may point to a partial conclusion, that in the onset-of-invasion period, the conflict attention is important in explaining day-ahead volatility throughout all of the indices. However, this may not hold for individual countries.

\begin{table}[h]
\centering
\setlength{\tabcolsep}{5.5pt} 
\renewcommand{\arraystretch}{1.2} 
\footnotesize 

\caption{Exploration of the significance of attention variables across all market indices}
\label{table:prelim-panels}

\begin{center}
\begin{tabular}{ l c c} 
    \hline
     & FE-DK-pre & FE-DK-during \\
    \hline
    $const$ & $\textbf{-2.572}^b$ & $\textbf{-3.091}^a$ \\
    
    $V_{i,t}$ & $\textbf{0.158}^d$ & $\textbf{0.196}^c$\\
    
    $V^w_{i,t}$ & $\textbf{0.326}^d$ & $0.167$ \\
    
    $C_{i,t}$ & $-0.056$ & $\textbf{0.245}^d$ \\
    
    $G_{i,t}$ & $\textbf{0.578}^b$ & $0.704$ \\
    
    \hline
    $R^2 (within) $ & 0.109 & 0.231 \\
    $R^2 (between) $ & 0.728 & 0.579 \\
    $F-statistic$ & 203.721 & 162.120 \\
    \hline
\end{tabular}
\end{center}

\textit{Notes: FE-DK-pre is the Equation \ref{eq:panel-har} estimated on the pre-invasion data sample, whereas FE-DK-during is estimated on the onset-of-invasion sample. Superscripts $a$, $b$, $c$ and $d$ denote statistical significance of estimated coefficients at the 10\%, 5\%, 1\% and 0.1\% level.}
\end{table}

Therefore, we estimate the model defined in Equation \ref{eq:har-model}
for all 51 stock market indices for both the before the invasion and
onset-of-invasion periods. The results show that in 70\% of countries,
the conflict attention variable $C_t$  has a significant positive
effect on future volatility. In other words, the more common military
conflict topic searches are, as measured by Google searches, the
higher the next day's volatility of MSCI indices. Table \ref{table1}
then presents estimates and diagnostics for those indices for which we
found the most significant impact of conflict attention, while Table
\ref{table2} shows the least significant impacts. The parameter
estimates of the remaining countries are reported in the Appendix in Table \ref{table3} and Table \ref{table4}. Each table also compares the results for the sample before the invasion of Ukraine (Panel B) and during the onset of the invasion (Panel A).

\begin{table}[h!]

\centering
\setlength{\tabcolsep}{2pt} 

\renewcommand{\arraystretch}{1.1} 
\scriptsize

\caption{Results with the most significant conflict attention variable $C_t$}
\label{table1}
\begin{center}
\begin{tabular}{ l c c c c c c c c c c } 

\hline
 & Poland & Denmark & Czechia & Great Britain & Portugal & Hungary & Belgium & Greece & Finland & Italy\\
 
\hline
\hline

\multicolumn{11}{l}{\textit{Panel A: OLS parameters estimates - with the most significant $C_t$ }} \\

Constant & 1.764 & $\textbf{-4.505}^a$ & -0.597 & -3.809 & -2.885 & 3.725 & $\textbf{-4.965}^a$ & 1.353 & -3.579 & -4.192 \\
$V_t$ & 0.033 & -0.093 & 0.252 & 0.129 & 0.200 & 0.139 & 0.040 & 0.013 & 0.057 & 0.150 \\
$V_w$ & $\textbf{-0.632}^d$ & $\textbf{-0.822}^c$ & -0.161 & 0.040 & $\textbf{-0.460}^b$ & 0.215 & -0.127 & -0.122 & 0.012 & -0.005 \\
$C_t$ & $\textbf{1.198}^d$ & $\textbf{0.715}^d$ & $\textbf{0.782}^d$ & $\textbf{0.504}^d$ & $\textbf{0.684}^d$ & $\textbf{0.703}^d$ & $\textbf{0.723}^d$ & $\textbf{0.653}^d$ & $\textbf{0.618}^d$ & $\textbf{0.652}^d$ \\
$G_t$  & -0.569 & $\textbf{1.329}^b$ & -0.230 & 0.654 & 0.677 & -1.036 & 1.038 & -0.515 & 0.801 & 0.876 \\

\hline
\multicolumn{11}{l}{\textit{Panel A1: Estimated models diagnostics}} \\

$R^2$ & 0.788 & 0.486 & 0.686 & 0.570 & 0.668 & 0.756 & 0.642 & 0.499 & 0.614 & 0.586 \\
$adj. R^2$ & 0.767 & 0.436 & 0.656 & 0.527 & 0.636 & 0.733 & 0.607 & 0.448 & 0.575 & 0.546 \\
residual $\rho(1)$ & 0.112 & 0.144 & 0.066 & 0.032 & 0.124 & -0.036 & 0.083 & 0.040 & 0.093 & 0.050 \\
Ljung-Box & 0.491 & 0.854 & 0.419 & 0.926 & 0.702 & 0.036 & 0.197 & 0.010 & 0.642 & 0.283 \\
White & 0.612 & 0.950 & 0.840 & 0.889 & 0.202 & 0.772 & 0.828 & 0.777 & 0.559 & 0.122 \\

\hline
\hline
\multicolumn{11}{l}{\textit{Panel B: Parameters estimates of the same model using pre-invasion data sample}} \\

Constant & $\textbf{-5.279}^c$ & $\textbf{-3.590}^b$ & $\textbf{-4.859}^c$ & -2.337 & $\textbf{-3.400}^a$ & -3.071 & -2.047 & -1.444 & $\textbf{-4.604}^b$ & -0.479 \\
$V_t$ & 0.030 & $\textbf{0.313}^d$ & $\textbf{0.395}^d$ & 0.123 & 0.085 & $\textbf{0.227}^b$ & 0.138 & 0.130 & 0.001 & $\textbf{0.262}^c$ \\
$V_w$  & $\textbf{0.541}^d$ & 0.147 & 0.079 & $\textbf{0.292}^a$ & 0.210 & $\textbf{0.365}^b$ & 0.163 & 0.029 & $\textbf{0.532}^d$ & 0.181 \\
$C_t$ & -0.173 & -0.313 & 0.217 & 0.067 & 0.075 & 0.057 & 0.098 & -0.208 & -0.240 & -0.028 \\
$G_t$  & $\textbf{1.311}^b$ & $\textbf{0.896}^b$ & $\textbf{1.142}^b$ & 0.388 & $\textbf{0.830}^a$ & 0.766 & 0.357 & 0.286 & $\textbf{1.087}^a$ & -0.016 \\

\hline
\multicolumn{11}{l}{\textit{Panel B1: Estimated models diagnostics}} \\

$R^2$ & 0.249 & 0.185 & 0.263 & 0.076 & 0.080 & 0.227 & 0.059 & 0.028 & 0.176 & 0.119 \\
$adj. R^2$ & 0.226 & 0.160 & 0.240 & 0.048 & 0.053 & 0.203 & 0.031 & -0.002 & 0.151 & 0.093 \\
residual $\rho(1)$ & 0.004 & -0.017 & -0.050 & 0.013 & 0.020 & -0.017 & 0.009 & 0.014 & -0.000 & -0.024 \\
Ljung-Box & 0.584 & 0.793 & 0.750 & 0.207 & 0.176 & 0.994 & 0.567 & 0.489 & 0.556 & 0.948 \\
White & 0.214 & 0.086 & 0.558 & 0.181 & 0.109 & 0.302 & 0.044 & 0.931 & 0.009 & 0.384 \\

\hline
\end{tabular}
\end{center}

\textit{Notes: In the header, we use ISO 3166 country codes. Regression parameter estimates in bold indicate significance at the 10\% level; superscripts $a$, $b$, $c$ and $d$ denote statistical significance of estimated coefficients at the 10\%, 5\%, 1\% and 0.1\% level. Residuals $\rho(1)$ describe the first-order autocorrelation of the residuals. For the Ljung-Box and White tests, we report the corresponding p-values.}
\end{table}
\begin{table}[h!]

\centering
\setlength{\tabcolsep}{2.5pt} 
\renewcommand{\arraystretch}{1.1} 
\scriptsize

\caption{Results with the least significant conflict attention variable $C_t$}
\label{table2}
\begin{center}
\begin{tabular}{ l c c c c c c c c c c } 

\hline
 & Brazil & Jordan & Argentina & Korea & Philippines & Taiwan & Japan & Columbia & Canada & Peru\\
 
\hline
\hline

\multicolumn{11}{l}{\textit{Panel A: OLS parameters estimates - with the least significant $C_t$ }} \\

Constant & $\textbf{-12.904}^a$ & -8.610 & $\textbf{-11.535}^c$ & -1.054 & 1.189 & $\textbf{-6.584}^b$ & 0.834 & -5.844 & $\textbf{-13.732}^d$ & $\textbf{-11.051}^c$ \\
$V_t$ & 0.115 & 0.131 & 0.191 & 0.110 & 0.267 & -0.065 & 0.090 & 0.002 & $\textbf{0.281}^a$ & 0.089 \\
$V_t^w$ & -1.110 & -0.155 & -0.152 & 0.187 & -0.166 & $\textbf{-0.824}^c$ & 0.211 & 0.360 & -0.071 & -0.001 \\
$C_t$& -0.784 & 0.175 & 0.106 & 0.096 & 0.116 & -0.066 & 0.087 & 0.075 & 0.021 & -0.012 \\
$G_t$ & $\textbf{3.898}^a$ & 1.990 & $\textbf{3.252}^d$ & 0.197 & -0.424 & $\textbf{1.599}^b$ & -0.265 & 1.493 & $\textbf{3.373}^d$ & $\textbf{3.018}^c$ \\

\hline
\multicolumn{11}{l}{\textit{Panel A1: Estimated models diagnostics}} \\

$R^2$ & 0.097 & 0.136 & 0.283 & 0.102 & 0.079 & 0.197 & 0.059 & 0.040 & 0.386 & 0.197 \\
$adj. R^2$ & 0.009 & 0.024 & 0.209 & 0.005 & -0.018 & 0.099 & -0.043 & -0.054 & 0.323 & 0.114 \\
residual $\rho(1)$ & -0.018 & 0.038 & 0.015 & 0.029 & 0.042 & -0.025 & 0.059 & -0.024 & 0.058 & 0.003 \\
Ljung-Box & 0.987 & 0.145 & 0.893 & 0.805 & 0.577 & 0.539 & 0.581 & 0.967 & 0.777 & 0.520 \\
White& 0.000 & 0.248 & 0.875 & 0.650 & 0.642 & 0.114 & 0.265 & 0.413 & 0.443 & 0.094 \\

\hline
\hline
\multicolumn{11}{l}{\textit{Panel B: Parameters estimates of the same model using pre-invasion data sample}} \\

Constant & $\textbf{-5.337}^a$ & 2.077 & -1.001 & $\textbf{-5.077}^c$ & -1.986 & 0.382 & $\textbf{-2.984}^a$ & -1.977 & $\textbf{-6.946}^c$ & 1.022 \\
$V_t$  & -0.077 & $\textbf{0.318}^c$ & $\textbf{0.272}^c$ & 0.109 & $\textbf{0.356}^d$ & 0.068 & 0.156 & 0.019 & $\textbf{0.217}^b$ & 0.170 \\
$V_t^w$ & $\textbf{0.587}^b$ & 0.018 & $\textbf{0.431}^c$ & $\textbf{0.536}^d$ & -0.017 & $\textbf{0.335}^a$ & 0.267 & 0.256 & $\textbf{0.281}^a$ & -0.058 \\
$C_t$ & -0.429 & -0.080 & -0.170 & $\textbf{-0.413}^a$ & 0.099 & 0.114 & -0.320 & -0.348 & -0.296 & 0.182 \\
$G_t$ & $\textbf{1.471}^a$ & -0.652 & 0.323 & $\textbf{1.243}^c$ & 0.429 & -0.238 & 0.657 & 0.513 & $\textbf{1.611}^c$ & -0.139 \\

\hline
\multicolumn{11}{l}{\textit{Panel B1: Estimated models diagnostics}} \\

$R^2$ & 0.092 & 0.114 & 0.270 & 0.295 & 0.133 & 0.069 & 0.091 & 0.018 & 0.183 & 0.029 \\
$adj. R^2$ & 0.065 & 0.079 & 0.248 & 0.273 & 0.106 & 0.040 & 0.062 & -0.012 & 0.157 & -0.001 \\
residual $\rho(1)$ & 0.009 & -0.071 & 0.033 & -0.020 & 0.025 & -0.018 & 0.009 & -0.001 & 0.018 & -0.014 \\
Ljung-Box & 0.981 & 0.401 & 0.632 & 0.970 & 0.853 & 0.429 & 0.230 & 1.000 & 0.919 & 0.244 \\
White & 0.797 & 0.476 & 0.909 & 0.961 & 0.768 & 0.121 & 0.922 & 0.958 & 0.088 & 0.042 \\

\hline
\end{tabular}
\end{center}

\textit{Notes: In the header, we use ISO 3166 country codes. Regression parameter estimates in bold indicate significance at the 10\% level; superscripts $a$, $b$, $c$ and $d$ denote statistical significance of estimated coefficients at the 10\%, 5\%, 1\% and 0.1\% level. Residuals $\rho(1)$ describe the first-order autocorrelation of the residuals. For the Ljung-Box and White tests, we report the corresponding p-values.}
\end{table}

In some cases the model diagnostics show mild heteroskedasticity and
autocorrelation of residuals as indicated by the p-values of the White
and Ljung-Box tests (see Tables \ref{table1}, \ref{table2},
\ref{table3} and \ref{table4}). To overcome these issues, we applied
the Newey-Further-West estimator \citep{10.2307/1913610}. We decided to apply
this method to all indices because this facilitates comparing the
resulting t-statistics across estimated models. We tested the
explanatory variables for the presence of a unit root via the test of
\citet{pesaran2007simple} for panel data stationarity. This test
rejected the null hypothesis of a unit root across the cross-section
of the volatility series. 
Some of the estimated models possess a low $R^2$ value, which may be a
result of a low persistence of the estimated volatility proxy in the pre-invasion sample. In models reported in Tables \ref{table1} and
\ref{table3}, the conflict attention measures essentially replace the
traditional role of daily volatility, which is apparent in the
substantially higher $R^2$ values than those observed in the before invasion period.

\begin{figure}[htp]
  \centering
  \caption{The significance and size of the impact the conflict
    attention index on future volatility at the onset-of-invasion -- worldwide} \label{map}
  \includegraphics[width=12.5cm]{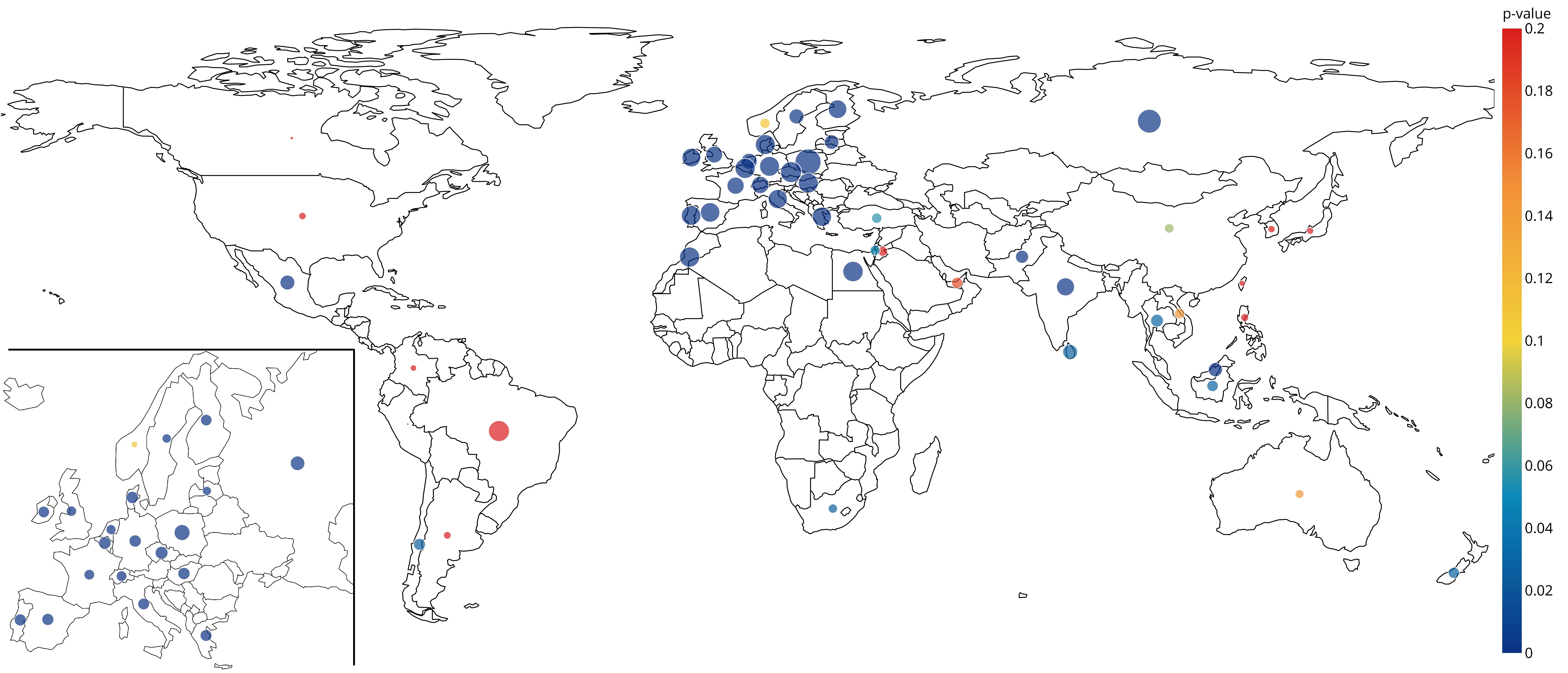}
  \fnote{Notes: The size of a circle is determined by the coefficient
    estimate of $C_t$, relative to the largest observed value. The
    color denotes the statistical significance in terms of the p-value
    of the $C_t$ coefficient estimate. With all p-values higher than
    0.2 being statistically insignificant, these values were capped at
    0.2 to facilitate graphical presentation.}
\end{figure}

Figure \ref{map} summarizes the results of our study. The dots mark
the countries analyzed, with their color determined by the p-value of
conflict attention variable $C_t$ and their size determined by the
magnitude of these parameter estimates. We can see that conflict
attention primarily affects the volatility in European countries,
where the conflict increases volatility. The countries outside of
Europe are rarely significantly affected and show a lower impact on their indices' volatility. 

\begin{figure}[h!]
  \centering
  \caption{The importance of conflict attention independence to economic connectedness with Russia and geographical distance} \label{chart}
\includegraphics[width=12.5cm]{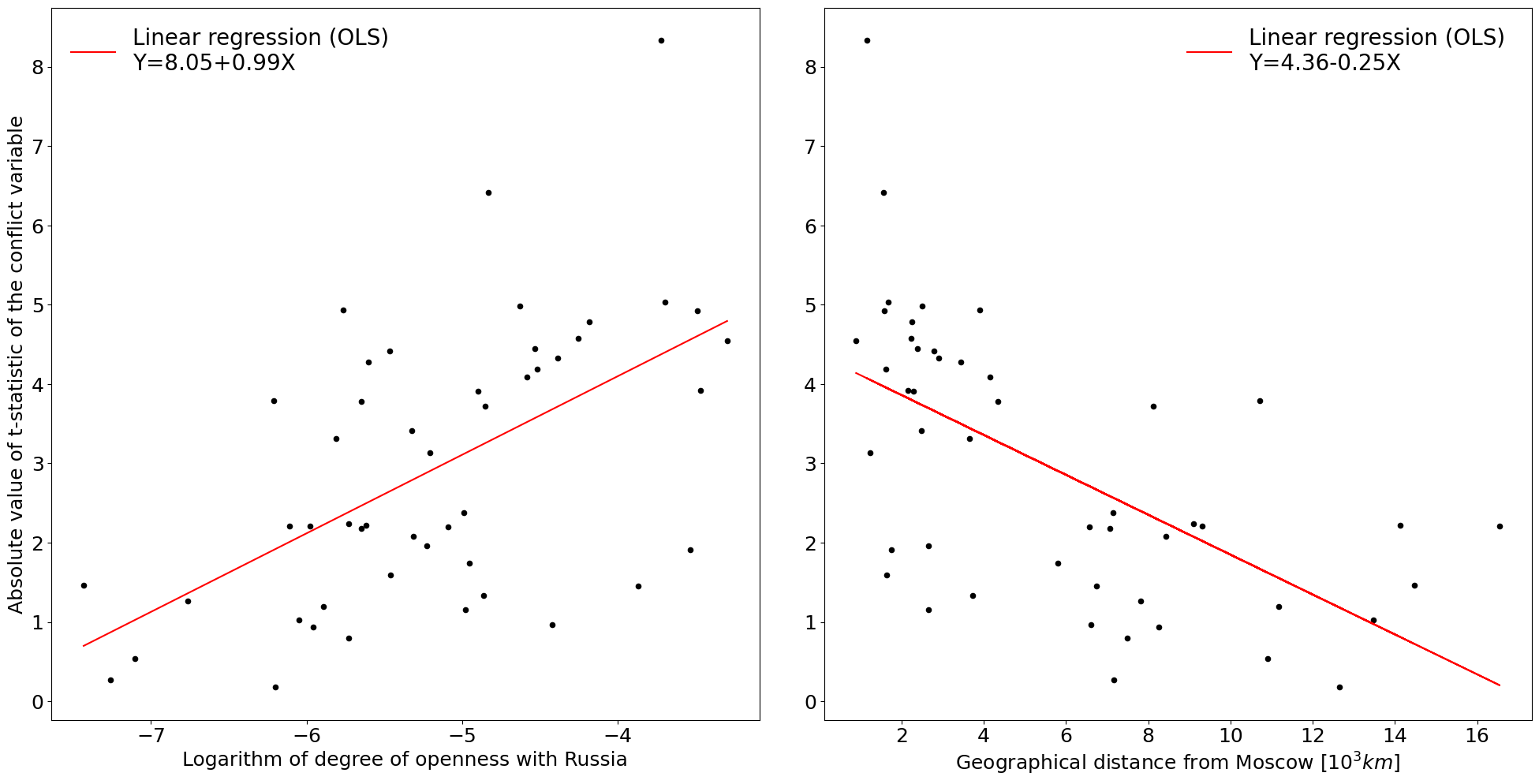}
\fnote{Notes: On the y-axis, we use the absolute value of the
  t-statistic for estimated parameter $C_t$ during the onset-of-invasion
  sample. We excluded Latvia from the chart for better graphical
  representation because its DOO value places it far from the other observations. However, this decision does not bias the presented results in any way.}
\end{figure}

Based on the results in Table \ref{table1}, the day-ahead volatility is most significantly affected by conflict attention at the onset of the invasion period in ten European countries. In addition, we find that all European countries in our dataset except for Norway are highly impacted. In fact, the effect is concentrated in Europe and near the military conflict, which is also visible in Figure \ref{map}. We assume that these countries might be influenced due to
strong economic ties with Russia, the threat of wider European conflict with the possible involvement of NATO or the ongoing humanitarian crisis (with a few million refugees fleeing Ukraine). Of the countries facing a wave of refugees, our analysis covers Poland, Czechia, Hungary, and Germany, all of which are among the most impacted countries. The previously mentioned dependence of Europe on oil and gas imports from Russia may also be an important reason why
this conflict has primarily affected European stock markets. This argument is particularly supported by the fact that we have not found a significant impact on Norway, which is independent of Russia due to the former's substantial oil and gas reserves. 

However, in the pre-invasion period, the conflict attention variable was not significant at the 5 \% level for any of the countries considered. This statement is also valid for wartime data for the countries extracted in Table \ref{table2}. These are primarily American and Asian countries. Regarding the variable representing the previous day's volatility $V_t$, despite the significance of this parameter for many countries in the before invasion period, $V_t$ became insignificant for most of the indices during the onset-of-invasion. 

There are also some significantly impacted countries outside of Europe, for instance, Mexico, whose results could be explained by the fact that Mexico did not condemn the Russian invasion \citep{ReutersMexico}. Various reasons could explain the impact of the conflict in other countries. We have, for example, countries that are dependent on wheat imports from Ukraine, or those, that like Russia, extract oil.

Based on these findings, we assume that the effect of the conflict attention variable increases the geographically closer the country is to Russia or the stronger its relations are with Russia. Firstly, we decided to graphically verify this relationship in Figure \ref{chart}, which displays the relationship between the conflict attention t-value of
each country and its economic openness with Russia and between the conflict attention t-value and its geographical distance from Moscow. 
Based on these charts, we conclude that the more open a country is to Russia, meaning a higher ratio of Russian imports and exports to the country's GDP, the more significant that conflict attention is. Conversely, conflict attention is less significant for countries that are more geographically distant from the Russian capital. \\
Secondly, appropriate tests were carried out. More specifically, we estimate the panel regression defined by Equation \ref{eq:panel-har-ext} for both variables $DOO_i$ and $Dist_i$. We are interested in the significance of the parameter $\beta_5$ as well as in its sign. Presented in Table \ref{table:panel-data-cd-cdoo}, the $\beta_5$ is significant in both panel regressions. The sign in case of the interaction term $C_{i,t}Dist_i$ is negative, and in case of $C_{i,t}DOO_i$ positive. This is in perfect alignment with Figure \ref{chart}. Simply put, marginal effect of a change in $Dist_i$ on day-ahead volatility is $\beta_5 C_{i,t}$, thus for an arbitrary $C_{i,t}$ (which is always positive) we are affected less by $C_{i,t}$ with the growing distance from Moscow. Analogically, the same interpretation holds for $DOO_i$, but with the opposite effect.

\begin{table}[h]
\centering
\setlength{\tabcolsep}{15.5pt} 
\renewcommand{\arraystretch}{1.2} 
\footnotesize 

\caption{Panel regression to provide further evidence on findings in Figure \ref{chart}}
\label{table:panel-data-cd-cdoo}

\begin{center}
\begin{tabular}{ l c c } 
    \hline
     & FE-DK-DOO & FE-DK-Dist \\
    \hline
    $const$ & $-3.125^a$ & $-3.280^b$ \\
    
    $V_{i,t}$ & $\textbf{0.193}^c$ & $\textbf{0.187}^c$ \\
    
    $V^w_{i,t}$ &  0.153 & $0.089$ \\
    
    $C_{i,t}$ & $\textbf{0.207}^c$ &  $\textbf{0.454}^d$ \\
    
    $G_{i,t}$ &  $0.712$ & $0.747^a$ \\
    
    $C_{i,t}DOO_{i}$ & $\textbf{3.731}^d$ \\
    
    $C_{i,t}Dist_{i}$ & & $\textbf{-0.033}^d$ \\
    \hline
    $R^2 (within)$ & 0.236 & 0.249 \\
    $R^2 (between)$ & 0.552 & 0.402 \\
    $F-statistic$ & 133.601 & 143.972 \\
    \hline
\end{tabular}
\end{center}

\textit{Notes: Superscripts $a$, $b$, $c$ and $d$ denote statistical significance of estimated coefficients at the 10\%, 5\%, 1\% and 0.1\% level.}
\end{table}

\subsection{Robustness checks of achieved results}

Achieved results and their significance may be influenced by several factors of choice we had to make in order to conduct our analysis. The most crucial ones are the time-period split date, log transformation of price variation variables, and the likeliness of the results being driven only by the first few days of the invasion. Thus, we resorted to checking our results under different settings. We tried three different split dates (2021-12-01, 2021-12-15, and 2022-01-15), under which the results point to the same conclusions -- the same hold for the no-log transformation of volatility measures. For controlling for the first week of the invasion, we introduce a dummy variable into model defined in Equation \ref{eq:har-model}. 

\begin{equation}
\label{eq:dummy-har-model}
V_{t+1} = \beta_0 + \beta_1 V_t + \beta_2 V_t^w + \beta_3 C_t +  \beta_4 D_t C_t + \beta_5 G_t + \epsilon_{t+1}
\end{equation}
 
where we assign $D_t$ with ones for the first week of invasion starting on 21st February and also for 21st and 24th February separately. The conflict variable remained significant.

Further, we estimate the Equation \ref{eq:dummy-sep-har-model}, where in contrast to Equation \ref{eq:dummy-har-model}, the effects of $D_t$ and $C_t$ are separated. 

\begin{equation}
\label{eq:dummy-sep-har-model}
V_{t+1} = \beta_0 + \beta_1 V_t + \beta_2 V_t^w + \beta_3 C_t +  \beta_4 D_t + \beta_5 G_t + \epsilon_{t+1}
\end{equation}

In this setup, we control for any external effect that may occur during the first week of the invasion, but is not contained by any of our independent variables. With the $\beta_4$ being mostly insignificant across the 51 countries. The only difference we observe is in case of Turkey and Israel, which $\beta_3$ became less significant, meaning there could be another factor we did not account for in these two countries. Results of other countries remains unchanged. 

Furthermore, in our main results, we employ the attention variables of international investors measured by worldwide Google Trends. To verify whether this attention was in fact international, we estimate our models again with country-specific conflict attention $C_t$ retrieved with location specified only for the country of the given MSCI index. Panel data regression with local conflict attention confirms our findings, as the results are nearly identical to those with international attention. On the other hand, for individual countries we find slight differences, which could be extended in future research.

\section{Conclusion}
\label{sec:conclusion}

This paper explores how investors' attention to the conflict between Russia and Ukraine influences the variability of asset prices in specific countries. We construct a Google search-based military conflict attention variable and general stock market attention to capture the effect of excessive attention devoted to the conflict. To draw sharp conclusions about volatility, we applied an HAR-RV model with a range-volatility estimator to MSCI data.  

Our results demonstrate that while the impact of the conflict
attention measure was insignificant in the pre-invasion period, at a time of escalating war threats, attention to conflict significantly affects volatility. Specifically, increasing conflict attention leads to higher volatility of the indices of the studied countries. The analysis of the indicators of the economic and geographical interconnectedness of individual countries to Russia shows that the effect of attention is more significant in countries with higher openness with Russia and those nearer to it. 

Future scope of the research may contain the study of volatility spillover effect among markets, as our results showed that for 36 countries, conflict attention variable is significant, which can be caused by contagion, similar to studies \citet{pericoli2003primer, diebold2009measuring, diebold2012better, Lyocsa2020a, okorie2021stock}. Our findings may also be applied in the context of portfolio asset allocation, which utilizes methods of Google Trends as \citet{Kristoufek2013, maggi2021google} or with adding the information regarding sentiment, the methods like \citet{dumas2009equilibrium, song2017stock, yu2022dynamic}. Finally, our findings may prove to be useful in the domain of option pricing as a better volatility estimate leads to a more appropriate option price \citep{Cretarola2017}.

\appendix


{\setstretch{1.0}
\begin{footnotesize}
\setlength{\bibsep}{0pt plus 0.3ex}
\Urlmuskip=0mu plus 1mu\relax
\bibliographystyle{elsarticle-harv} 
\bibliography{cas-refs}
\end{footnotesize}}




\section{Appendix}

The list of general attention topics is as follows: 'asset allocation', 'Bloomberg', 'day trading', 'dividend yield', 'earnings call', 'earnings per share', 'exchange-traded fund', 'financial crisis', 'financial market', 'futures contract', 'Google Finance', 'government bond', 'hedge fund', 
'Implied volatility', 'market capitalization', 'market liquidity', 'market sentiment', 'MSCI', 'mutual fund', 'option contract', "pension fund', 'price–earnings ratio', 'quarterly finance report', 'stock market index', 'stock market', 'technical analysis', 'ticker symbol', 'VIX', 'volatility', 'Yahoo! Finance', and 'yield curve'.
\begin{landscape}
\label{sec:appendix}
\begin{table}[h!]

\centering
\setlength{\tabcolsep}{1pt} 
\renewcommand{\arraystretch}{1.1} 
\small

\caption{Further results with significant conflict attention variable $C_t$}
\label{table3}
\begin{center}
\begin{tabular}{ l c c c c c c c c c c c c c c c} 

\hline
  & IRL & EGY & ESP & DEU & MAR & NLD & CHE & MEX & IND & MYS & FRA & LVA & PAK & SWE & RUS\\
 
\hline
\hline

\multicolumn{16}{l}{\textit{Panel A: OLS parameters estimates -- during war period }} \\

Constant & -5.577 & -1.471 & -5.929 & -2.858 & 4.633 & -1.449 & -3.285 & $\textbf{-9.526}^c$ & $\textbf{-9.169}^b$ & 0.210 & -2.308 & $\textbf{-8.870}^a$ & -3.000 & -0.240 & 7.026 \\
$V_t$ & 0.011 & $\textbf{0.313}^a$ & 0.152 & 0.081 & 0.055 & 0.099 & 0.127 & 0.158 & 0.033 & -0.010 & $\textbf{0.265}^a$ & -0.039 & 0.235 & 0.168 & 0.122  \\
$V_t^w$& -0.098 & -0.462 & -0.191 & -0.131 & 0.014 & -0.397 & -0.301 & -0.182 & -0.147 & 0.232 & -0.057 & 0.731 & $\textbf{-0.636}^b$ & 0.031 & 0.121  \\
$C_t$ & $\textbf{0.625}^d$ & $\textbf{0.740}^d$ & $\textbf{0.679}^d$ & $\textbf{0.707}^d$ & $\textbf{0.725}^d$ & $\textbf{0.466}^d$ & $\textbf{0.535}^d$ & $\textbf{0.396}^d$ & $\textbf{0.582}^d$ & $\textbf{0.362}^d$ & $\textbf{0.527}^d$ & $\textbf{0.380}^d$ & $\textbf{0.309}^d$ & $\textbf{0.406}^c$ & $\textbf{1.006}^c$  \\
$G_t$ & 1.389 & 0.070 & 1.325 & 0.490 & -1.882 & 0.509 & 0.621 & $\textbf{2.344}^c$ & $\textbf{2.116}^b$ & -0.473 & 0.429 & $\textbf{2.066}^a$ & 0.628 & 0.033 & -1.670 \\

\hline
\multicolumn{16}{l}{\textit{Panel A1: Estimated models diagnostics}} \\

$R^2$ & 0.502 & 0.438 & 0.554 & 0.570 & 0.466 & 0.359 & 0.425 & 0.556 & 0.536 & 0.412 & 0.620 & 0.494 & 0.262 & 0.436 & 0.669 \\
$adj. R^2$ & 0.452 & 0.364 & 0.510 & 0.528 & 0.413 & 0.297 & 0.368 & 0.513 & 0.488 & 0.350 & 0.583 & 0.444 & 0.190 & 0.379 & 0.636 \\
residual $\rho(1)$ & 0.120 & 0.198 & 0.022 & 0.089 & -0.001 & 0.079 & 0.072 & -0.045 & 0.110 & -0.110 & 0.086 & 0.038 & 0.004 & 0.046 & 0.076 \\
Ljung-Box & 0.691 & 0.514 & 0.133 & 0.907 & 0.861 & 0.917 & 0.331 & 0.017 & 0.060 & 0.775 & 0.877 & 0.902 & 0.256 & 0.985 & 0.469 \\
White& 0.287 & 0.780 & 0.967 & 0.061 & 0.668 & 0.104 & 0.102 & 0.947 & 0.399 & 0.338 & 0.454 & 0.001 & 0.500 & 0.459 & 0.690 \\

\hline
\hline
\multicolumn{16}{l}{\textit{Panel B: Parameters estimates of the same model using pre-invasion data sample}} \\

Constant & $\textbf{-3.524}^b$ & -3.362 & -2.815 & -0.652 & -0.040 & -3.825 & -1.159 & $\textbf{-5.427}^d$ & -2.202 & -1.898 & -3.393 & $\textbf{11.049}^a$ & $\textbf{-3.295}^a$ & -1.160 & $\textbf{-3.363}^b$ \\
$V_t$ & 0.113 & $\textbf{0.213}^a$ & 0.147 & $\textbf{0.207}^b$ & 0.115 & $\textbf{0.174}^a$ & 0.170 & $\textbf{0.181}^a$ & $\textbf{0.308}^c$ & 0.065 & 0.094 & 0.062 & $\textbf{0.200}^b$ & $\textbf{0.345}^d$ & 0.079 \\
$V_t^w$ & $\textbf{0.311}^b$ & 0.184 & 0.339 & 0.225 & 0.301 & $\textbf{0.441}^c$ & $\textbf{0.319}^b$ & 0.212 & 0.228 & 0.250 & $\textbf{0.509}^c$ & 0.197 & $\textbf{0.483}^d$ & $\textbf{0.288}^a$ & $\textbf{0.545}^d$  \\
$C_t$ & $\textbf{-0.496}^a$ & -0.075 & -0.107 & 0.141 & -0.296 & -0.251 & 0.151 & -0.117 & -0.063 & -0.061 & -0.154 & $\textbf{0.697}^a$ & $\textbf{-0.402}^a$ & -0.060 & 0.183 \\
$G_t$ & $\textbf{0.875}^a$ & 0.898 & 0.662 & -0.029 & -0.310 & 0.939 & 0.127 & $\textbf{1.385}^d$ & 0.463 & 0.205 & 0.739 & $\textbf{-3.142}^b$ & $\textbf{0.863}^a$ & 0.223 & $\textbf{0.864}^b$  \\

\hline
\multicolumn{16}{l}{\textit{Panel B1: Estimated models diagnostics}} \\

$R^2$  & 0.104 & 0.105 & 0.123 & 0.104 & 0.072 & 0.206 & 0.154 & 0.170 & 0.200 & 0.032 & 0.177 & 0.048 & 0.309 & 0.281 & 0.306 \\
$adj. R^2$ & 0.076 & 0.068 & 0.096 & 0.077 & 0.043 & 0.182 & 0.128 & 0.145 & 0.175 & 0.002 & 0.153 & 0.018 & 0.288 & 0.259 & 0.285 \\
residual $\rho(1)$ & 0.007 & 0.008 & 0.008 & 0.010 & 0.003 & -0.005 & -0.016 & 0.043 & 0.030 & -0.009 & 0.006 & -0.008 & -0.065 & -0.061 & 0.006\\
Ljung-Box & 0.132 & 0.998 & 0.610 & 0.908 & 0.038 & 0.622 & 0.792 & 0.880 & 0.113 & 0.733 & 0.597 & 0.909 & 0.238 & 0.470 & 0.013 \\
White & 0.320 & 0.420 & 0.000 & 0.275 & 0.924 & 0.031 & 0.005 & 0.429 & 0.786 & 0.008 & 0.000 & 1.000 & 0.356 & 0.005 & 0.009 \\

\hline
\end{tabular}
\end{center}

\textit{Notes: In the header, we use ISO 3166 country codes. Regression parameter estimates in bold indicate significance at the 10\% level; superscripts $a$, $b$, $c$ and $d$ denote statistical significance of estimated coefficients at the 10\%, 5\%, 1\% and 0.1\% level. Residuals $\rho(1)$ describe the first-order autocorrelation of the residuals. For the Ljung-Box and White tests, we report the corresponding p-values.}
\end{table}
\end{landscape}
\begin{landscape}
\begin{table}[h!]

\centering
\setlength{\tabcolsep}{1pt} 
\renewcommand{\arraystretch}{1.1} 
\small

\caption{Further results with less- or in-significant conflict attention variable $C_t$}
\label{table4}
\begin{center}
\begin{tabular}{ l c c c c c c c c c c c c c c c c} 

\hline
& HKG & ZAF & CHL & NZL & IDN & LKA & THA & SGP & ISR & TUR & CHN & NOR & AUS & VNM & ARE & USA\\
\hline
\hline

\multicolumn{17}{l}{\textit{Panel A: OLS parameters estimates -- during war }} \\

Constant & -5.332 & $\textbf{-4.269}^a$ & $\textbf{-5.595}^b$ & -3.369 & -3.751 & -3.274 & $\textbf{-4.999}^a$ & $\textbf{-8.351}^b$ & $\textbf{-12.086}^c$ & -3.442 & -2.869 & $\textbf{-6.179}^b$ & $\textbf{-9.089}^b$ & -3.183 & 0.202 & $\textbf{-13.040}^c$ \\
$V_t$ & $\textbf{0.474}^b$ & -0.180 & -0.113 & 0.073 & -0.070 & 0.045 & -0.215 & 0.238 & $\textbf{-0.350}^b$ & 0.044 & $\textbf{0.562}^c$ & 0.095 & 0.278 & 0.011 & 0.128 & $\textbf{0.294}^b$ \\
$V_t^w$ & $\textbf{-1.878}^a$ & $\textbf{0.529}^a$ & -0.493 & 0.284 & -0.011 & 0.123 & 0.373 & -0.150 & -0.114 & -0.414 & $\textbf{-0.703}^a$ & 0.140 & -0.082 & 0.461 & 0.061 & -0.066 \\
$C_t$ & $\textbf{0.635}^b$ & $\textbf{0.157}^b$ & $\textbf{0.258}^b$ & $\textbf{0.229}^b$ & $\textbf{0.235}^b$ & $\textbf{0.398}^b$ & $\textbf{0.294}^b$ & $\textbf{0.250}^b$ & $\textbf{0.182}^a$ & $\textbf{0.192}^a$ & $\textbf{0.165}^a$ & 0.189 & 0.141 & -0.197 & 0.226 & 0.099 \\
$G_t$ & 0.769 & 0.970 & $\textbf{1.677}^b$ & 0.747 & 0.783 & 0.984 & 0.939 & $\textbf{1.961}^b$ & $\textbf{3.227}^d$ & 1.172 & 0.663 & $\textbf{1.474}^a$ & $\textbf{2.129}^b$ & 0.954 & -0.283 & $\textbf{3.314}^c$ \\

\hline
\multicolumn{17}{l}{\textit{Panel A1: Estimated models diagnostics}} \\

$R^2$ & 0.368 & 0.283 & 0.246 & 0.337 & 0.200 & 0.301 & 0.326 & 0.331 & 0.340 & 0.159 & 0.307 & 0.282 & 0.308 & 0.269 & 0.115 & 0.353 \\
$adj. R^2$ & 0.306 & 0.213 & 0.172 & 0.268 & 0.116 & 0.223 & 0.257 & 0.266 & 0.276 & 0.077 & 0.240 & 0.212 & 0.236 & 0.185 & 0.028 & 0.286 \\
residual $\rho(1)$ & 0.081 & 0.023 & -0.021 & -0.064 & 0.037 & -0.004 & 0.093 & 0.021 & 0.071 & -0.027 & -0.146 & -0.025 & -0.059 & 0.016 & -0.033 & -0.073 \\
Ljung-Box & 0.976 & 0.893 & 0.920 & 0.984 & 0.825 & 0.001 & 0.827 & 0.239 & 0.109 & 0.714 & 0.018 & 0.468 & 0.832 & 0.645 & 0.787 & 0.572 \\
White & 0.237 & 0.357 & 0.157 & 0.535 & 0.653 & 0.794 & 0.026 & 0.077 & 0.287 & 0.786 & 0.674 & 0.123 & 0.767 & 0.654 & 0.309 & 0.495 \\

\hline
\hline
\multicolumn{17}{l}{\textit{Panel B: Parameters estimates of the same model using pre-invasion data sample}} \\

Constant & -2.657 & -1.966 & -2.037 & -2.477 & $\textbf{-3.705}^b$ & $\textbf{-3.348}^a$ & -2.054 & $\textbf{-4.134}^a$ & $\textbf{-13.124}^c$ & -1.489 & -1.722 & $\textbf{-4.839}^b$ & -2.646 & $\textbf{3.486}^b$ & -0.483 & -1.707 \\
$V_t$ & 0.176 & 0.125 & 0.104 & 0.077 & 0.076 & $\textbf{0.202}^a$ & 0.144 & 0.135 & -0.001 & $\textbf{0.237}^a$ & $\textbf{0.238}^b$ & 0.113 & $\textbf{0.315}^d$ & $\textbf{0.183}^a$ & 0.170 & $\textbf{0.425}^d$ \\
$V_t^w$ & 0.191 & 0.181 & 0.140 & $\textbf{0.338}^a$ & 0.225 & $\textbf{0.335}^b$ & $\textbf{0.359}^b$ & 0.134 & $\textbf{0.576}^b$ & $\textbf{0.599}^d$ & 0.263 & $\textbf{0.288}^a$ & -0.136 & $\textbf{0.358}^b$ & $\textbf{0.228}^a$ & 0.158 \\
$C_t$ & -0.169 & -0.342 & 0.109 & -0.084 & -0.200 & 0.208 & $\textbf{-0.506}^c$ & 0.428 & 0.164 & 0.188 & 0.061 & 0.198 & -0.039 & $\textbf{0.455}^b$ & $\textbf{0.603}^a$ & 0.162 \\
$G_t$ & 0.533 & 0.441 & 0.614 & 0.524 & $\textbf{0.877}^b$ & $\textbf{0.837}^a$ & 0.409 & 0.888 & $\textbf{3.282}^b$ & 0.370 & 0.415 & $\textbf{1.124}^b$ & 0.413 & $\textbf{-0.886}^b$ & -0.055 & 0.279 \\

\hline
\multicolumn{17}{l}{\textit{Panel B1: Estimated models diagnostics}} \\

$R^2$ & 0.057 & 0.061 & 0.026 & 0.070 & 0.072 & 0.228 & 0.127 & 0.100 & 0.249 & 0.550 & 0.131 & 0.161 & 0.096 & 0.267 & 0.170 & 0.276 \\
$adj. R^2$ & 0.029 & 0.031 & -0.003 & 0.041 & 0.044 & 0.204 & 0.099 & 0.073 & 0.227 & 0.535 & 0.105 & 0.135 & 0.068 & 0.245 & 0.136 & 0.253 \\
residual $\rho(1)$ & 0.014 & -0.007 & -0.003 & -0.011 & -0.020 & -0.025 & -0.020 & 0.023 & 0.046 & -0.023 & -0.004 & 0.018 & 0.024 & 0.004 & 0.022 & -0.080 \\
Ljung-Box & 0.679 & 0.425 & 0.994 & 0.132 & 0.225 & 0.561 & 0.461 & 0.167 & 0.801 & 0.152 & 0.800 & 0.646 & 0.592 & 0.884 & 0.423 & 0.481 \\
White & 0.782 & 0.000 & 0.759 & 0.030 & 0.840 & 0.458 & 0.121 & 0.086 & 0.000 & 0.103 & 0.702 & 0.000 & 0.299 & 0.832 & 0.832 & 0.703 \\

\hline
\end{tabular}
\end{center}

\textit{Notes: In the header, we use ISO 3166 country codes. Regression parameter estimates in bold indicate significance at the 10\% level; superscripts $a$, $b$, $c$ and $d$ denote statistical significance of estimated coefficients at the 10\%, 5\%, 1\% and 0.1\% level. Residuals $\rho(1)$ describe the first-order autocorrelation of the residuals. For the Ljung-Box and White tests, we report the corresponding p-values.}
\end{table}
\end{landscape}
\begin{table}[ht]

\centering
\setlength{\tabcolsep}{2.5pt} 
\renewcommand{\arraystretch}{1.1} 
\small

\caption{Descriptive statistics of daily price variation estimates $V_t$}
\label{daily-vol-descriptives}
\begin{center}

\begin{tabular}{ l c c c c c c c } 

\hline
Country & Mean & S.D. & Median & Min. & Max. & $\rho(1)$ & $\rho(5)$ \\
\hline

MEX & 1.386 & 1.573 & 0.952 & 0.080 & 14.892 & 0.297 & 0.096 \\
TWN & 0.742 & 0.694 & 0.545 & 0.047 & 4.935 & 0.122 & 0.006 \\
VNM & 2.374 & 2.949 & 1.550 & 0.140 & 21.005 & 0.336 & 0.206 \\
LVA & 2.094 & 13.852 & 0.436 & 0.000 & 178.848 & 0.224 & 0.280 \\
POL & 1.903 & 4.026 & 0.821 & 0.098 & 40.368 & 0.596 & 0.244 \\
NLD & 1.747 & 2.485 & 0.962 & 0.052 & 19.410 & 0.530 & 0.288 \\
ESP & 1.442 & 2.431 & 0.755 & 0.062 & 24.491 & 0.277 & 0.205 \\
JOR & 1.193 & 2.269 & 0.478 & 0.033 & 16.052 & 0.051 & -0.098 \\
CHL & 3.335 & 8.850 & 1.864 & 0.000 & 104.034 & 0.032 & -0.005 \\
ARE & 0.865 & 2.105 & 0.389 & 0.047 & 19.848 & 0.116 & -0.021 \\
GRC & 1.217 & 1.681 & 0.752 & 0.111 & 12.780 & 0.283 & 0.198 \\
ISR & 1.182 & 2.083 & 0.678 & 0.001 & 19.732 & 0.146 & 0.219 \\
JPN & 0.777 & 0.677 & 0.580 & 0.077 & 4.000 & 0.299 & 0.192 \\
SGP & 0.686 & 0.704 & 0.487 & 0.004 & 5.379 & 0.330 & 0.101 \\
EGY & 1.979 & 2.477 & 1.272 & 0.105 & 15.377 & 0.255 & 0.285 \\
MYS & 0.296 & 0.267 & 0.214 & 0.034 & 1.841 & 0.245 & 0.084 \\
FIN & 1.439 & 2.700 & 0.664 & 0.049 & 24.208 & 0.399 & 0.410 \\
BRA & 3.420 & 2.793 & 2.647 & 0.000 & 19.316 & 0.380 & -0.052 \\
RUS & 24.636 & 119.963 & 1.350 & 0.149 & 1305.617 & 0.294 & 0.244 \\
KOR & 0.742 & 0.687 & 0.510 & 0.083 & 5.219 & 0.350 & 0.065 \\
PER & 2.664 & 3.056 & 1.851 & 0.274 & 23.346 & 0.233 & -0.076 \\
PRT & 1.514 & 2.043 & 0.965 & 0.117 & 19.941 & 0.403 & 0.240 \\
CAN & 0.531 & 0.833 & 0.266 & 0.030 & 7.395 & 0.501 & 0.031 \\
DEU & 0.896 & 1.602 & 0.364 & 0.027 & 14.705 & 0.429 & 0.446 \\
CHE & 0.754 & 1.225 & 0.364 & 0.042 & 9.587 & 0.410 & 0.150 \\
IRL & 1.948 & 3.835 & 0.954 & 0.100 & 42.090 & 0.475 & 0.213 \\
MAR & 0.275 & 0.522 & 0.116 & 0.009 & 4.389 & 0.361 & 0.138 \\
IND & 0.905 & 1.295 & 0.524 & 0.061 & 12.761 & 0.330 & 0.197 \\
HKG & 0.700 & 0.811 & 0.477 & 0.001 & 7.215 & 0.178 & -0.012 \\
AUS & 0.459 & 0.570 & 0.311 & 0.057 & 6.195 & 0.393 & -0.011 \\
CZE & 0.982 & 1.864 & 0.425 & 0.036 & 19.126 & 0.435 & 0.323 \\
BEL & 1.062 & 2.139 & 0.458 & 0.076 & 18.469 & 0.349 & 0.237 \\
IDN & 0.866 & 0.692 & 0.686 & 0.079 & 4.826 & 0.150 & 0.064 \\
PHL & 0.948 & 1.061 & 0.610 & 0.048 & 8.059 & 0.292 & -0.081 \\
ITA & 1.180 & 2.553 & 0.510 & 0.047 & 28.591 & 0.504 & 0.496 \\
CHN & 1.244 & 1.117 & 0.923 & 0.039 & 8.078 & 0.445 & 0.131 \\
TUR & 3.252 & 8.524 & 1.250 & 0.145 & 77.889 & 0.659 & 0.089 \\
NZL & 0.793 & 0.787 & 0.561 & 0.076 & 6.595 & 0.331 & 0.171 \\
USA & 0.840 & 1.375 & 0.406 & 0.006 & 12.228 & 0.460 & 0.117 \\
ARG & 4.824 & 5.258 & 3.072 & 0.305 & 35.770 & 0.491 & 0.228 \\
DNK & 1.788 & 2.995 & 0.955 & 0.103 & 26.473 & 0.164 & 0.092 \\
FRA & 0.976 & 1.723 & 0.440 & 0.036 & 16.661 & 0.507 & 0.452 \\
COL & 1.994 & 2.211 & 1.284 & 0.001 & 17.012 & 0.247 & -0.015 \\
SWE & 1.331 & 2.323 & 0.589 & 0.066 & 22.986 & 0.456 & 0.254 \\
GBR & 0.544 & 0.694 & 0.304 & 0.028 & 5.649 & 0.392 & 0.292 \\
HUN & 3.392 & 9.141 & 1.072 & 0.110 & 70.508 & 0.731 & 0.511 \\
PAK & 1.722 & 1.717 & 1.180 & 0.178 & 12.174 & 0.442 & 0.186 \\
THA & 0.502 & 0.491 & 0.388 & 0.102 & 4.421 & 0.454 & 0.031 \\
ZAF & 0.868 & 0.660 & 0.647 & 0.067 & 4.874 & 0.197 & -0.001 \\
LKA & 3.035 & 5.642 & 1.064 & 0.084 & 42.756 & 0.567 & 0.302 \\
NOR & 0.860 & 1.014 & 0.497 & 0.045 & 9.059 & 0.214 & 0.123 \\

\hline

\end{tabular}
\end{center}

\textit{Notes: Countries are denoted by ISO 3166 country
  codes. S.D. stands for standard deviation, and $\rho(n)$ is an
  autocorrelation coefficient of the $n$th order.}
\end{table}
\end{document}